\documentclass[draftcls,onecolumn,12pt]{IEEEtran} 

\usepackage{cite}     
\usepackage{amsmath}    	
\usepackage{graphicx}  	
\usepackage{amssymb}
\usepackage{array}
\usepackage{stfloats,cancel}
\usepackage{balance}	
\usepackage{dsfont}
\usepackage{bbm}
\usepackage{ifpdf,xfrac}
\usepackage{stfloats,color}
\usepackage[T1]{fontenc}
\graphicspath{{./Figuras/}}

\newcommand{\figsize}{0.5}

\newtheorem{Lemma}{Lemma}[section]
\newtheorem{Theorem}{Theorem}[section]

\newtheorem{definition}{Definition}[section]
\newcommand {\be}{\begin{equation}}
\newcommand {\ee}{\end{equation}}

\newcommand{\prob}{\mathbb{P}}
\newcommand{\adf}{\mathcal{A}_{DF}}
\newcommand{\bdf}{\mathcal{B}_{DF}}
\newcommand{\acf}{\mathcal{A}_{CF}}
\newcommand{\bcf}{\mathcal{B}_{CF}}
\newcommand{\adt}{\mathcal{A}_{DT}}

\newcommand{\Lap}{\mathcal{L}}

\newcommand{\poutdf}{\prob_{\text{out,DF}}}
\newcommand{\poutdt}{\prob_{\text{out,DT}}}
\newcommand{\poutcf}{\prob_{\text{out,CF}}}

\newcommand{\Ex}{\mathbb{E}}
\newcommand{\R}{\mathbb{R}}

\newcommand{\Onot}{O}

\newcommand{\Cfunc}{\mathcal{C}}
\newcommand{\hsr}{h_{sr}}
\newcommand{\hsd}{h_{sd}}
\newcommand{\hrd}{h_{rd}}
\newcommand{\asdf}{\mathcal{A}_{SDF}}
\newcommand{\bsdf}{\mathcal{B}_{SDF}}

\newcommand{\Cn}{\mathcal{C}_n}

\begin{document}
\title{On the Outage Probability of the Full-Duplex Interference-Limited Relay Channel}

\author{Andr\'es Altieri, Leonardo Rey Vega, Pablo Piantanida, 
        and Cecilia G. Galarza
\thanks{This work was partially supported by DIGITEO-DIM  No. 2010-33D (ACRON), the Peruilh grant of the UBA and project UBACYT 2002010200250. The material in this paper was presented in part at the IEEE International Symposium on Information Theory, 2011.}
\thanks{A. Altieri (e-mail: aaltieri@fi.uba.ar, andres.altieri@supelec.fr). L. Rey Vega and C. Galarza are with the Departments of Electronics (FIUBA) and CSC-CONICET, Buenos Aires, Argentina (e-mail: lrey@fi.uba.ar, cgalar@fi.uba.ar).}
\thanks{P. Piantanida is with the Department of Telecommunications, SUPELEC, 91192 Gif-sur-Yvette, France (e-mail: pablo.piantanida@supelec.fr).}}

%


\maketitle

\begin{abstract}
In this paper, we study the performance, in terms of the asymptotic error probability, of a user which  communicates with a destination with the aid of a full-duplex in-band relay. We consider that the network is interference-limited, and interfering users are distributed as a Poisson point process. In this case, the asymptotic error probability is upper bounded by the outage probability (OP). We investigate the outage behavior for well-known cooperative schemes, namely, decode-and-forward (DF) and compress-and-forward (CF) considering fading and path loss. For DF we determine the exact OP and develop upper bounds which are tight in typical operating conditions. Also, we find the correlation coefficient between source and relay signals which minimizes the OP when the density of interferers is small. For CF, the achievable rates are determined by the spatial correlation of the interferences, and a straightforward analysis isn't  possible. To handle this issue, we show the rate with correlated noises is at most one bit worse than with uncorrelated noises, and thus find an upper bound on the performance of CF. These results are useful to evaluate the performance and to optimize relaying schemes in the context of full-duplex wireless networks.
\end{abstract}


%
\IEEEpeerreviewmaketitle

\section{Introduction}
\label{sec:intro}
In recent years, diversity-exploiting techniques for cooperative communications in wireless networks have been one of the most promising techniques to cope with the always increasing traffic demands. As such, strategies involving relays have received much attention as a means for improving the throughput and reliability of individual links \cite{kramer_cooperative_2005} \cite{laneman_cooperative_2004} \cite{host-madsen_capacity_2005}. For example, in the context of cellular networks, fourth-generation (4G) mobile-broadband systems allow for new coordination and cooperation strategies among base stations and relaying nodes. Basic relaying functionality was included in the Long Term Evolution (LTE) Rel. 10 standard \cite{PFD_2011}, while  Rel. 11 introduces several  Coordinated Multi-Point Operation (CoMP) modes\cite{RYS-WP2013}. In the former, communications can be established via a half-duplex relay node, wirelessly connected to a BS, either in the same frequency as the relay-destination link (in-band relaying), or in another band (out-of-band relaying) \cite{PFD_2011}. The latter considers homogeneous, i.e. single-tier, networks, with relays, and heterogeneous ones, where macro-cells and smaller cells are jointly deployed\cite{RYS-WP2013}.

Since the seminal paper by Cover and El Gamal \cite{cover_capacity_1979}, the relay channel has received much attention
from an information-theoretic perspective. In \cite{cover_capacity_1979}, the main coding strategies, decode-and-forward (DF) and compress-and-forward (CF) were introduced and analyzed. A third alternative is amplify-and-forward (AF) which we will not address in this paper. In general, works on additive-noise relay channels assume that transmissions are impaired by uncorrelated background noises at the receivers. Although this assumption has proved to be very useful, in the context of wireless networks, it may also be interesting to consider the interference correlation, which arises, among other things, because  many receivers will experience interference from the same sources. This correlation appears both in the interference time signals, and in the random, correlated interference powers at the receivers. Moreover, in a large network, users may interact, causing each other adverse interference conditions, an effect which is not present when uncorrelated background noises are considered.
Stochastic geometry \cite{stochastic_geometry2009} models have 
emerged as a useful and versatile toolbox for the analysis of large wireless networks, in which the interference among neighboring nodes is a key limit of performance \cite{BB2010,haenggi_stochastic_2009}. Through these models, the  random spatial distribution of the nodes,  random interference-signal correlation and powers,  and user-interaction, 
can be modeled in an elegant and compact fashion, leading to insightful results and guidelines in network design and analysis.  

In this paper, we study the performance, in terms of the asymptotic error probability, of full-duplex in-band relaying in a network impaired by interference between neighboring nodes, using an information-theoretic and stochastic geometry approach. 
Specifically, we consider the case in which a source attempts to communicate to a destination using the full-duplex DF or CF protocols. The gains in asymptotic error probability are upper bounded in terms of the \emph{outage probability} (OP), that is, the probability that, due to instantaneous conditions, the channel cannot support the rate attempted by a transmitting user. The co-channel interference experienced by the users is modeled using a homogeneous Poisson point process, considering both path loss attenuation and Rayleigh fading. 
\vspace{-3mm}
\subsection{Related work}
In the context of outage in slow-fading Gaussian networks, works generally consider that fixed-power, uncorrelated noises are present at the receivers, and perform analysis for each value of the noise powers, without considering correlated or random interference signals. For example,  \cite{katz_cooperative_2009} studies, among other things, the performance of  DF  in a single-relay network in which the source-relay link has a fixed, known amplitude, and the source does not know if the relay is present or not. The expected throughput of the scheme and the optimal correlation between the transmissions of the source and the relay are characterized for each value of the noise powers. In \cite{ZCL_2008}, a single-relay model in a Rayleigh fading environment is considered, with fixed, uncorrelated Gaussian background noise. The OP and the ergodic rate for DF and the correlation coefficient between the source and the relay, along with power  allocation between transmitters are considered. Some examples in which correlation is considered between the nodes are \cite{Goldsmith2011} and \cite{SSPK_2009}. Reference \cite{Goldsmith2011} studies the impact of noise correlation on the achievable rates of DF and CF for fixed-power background noises, while \cite{SSPK_2009} provides analytical expressions for the end-to-end SNR and OP of cooperative diversity in correlated lognormal channels for full-duplex DF relaying, and selection combining or maximum ratio combining, as a function of the joint channel densities.

There has been much work in the context of wireless networks with the aid of stochastic geometry models. The simplest of these models, the one which has been most frequently used is the homogeneous Poisson point process. The fundamental benefit of this process is that it generally leads to mathematically simpler and more tractable expressions than other, more structured processes. Initially, the Poisson process model for node distribution was taken to be a reasonable approximation for decentralized networks in which transmissions take place in uncoordinated fashion, such as with the ALOHA medium access strategy. Examples of these are wireless sensor and \emph{ad hoc} networks  \cite{baccelli_aloha_2006,weber_overview_2010}. In recent years, it has also been adopted for more structured networks such as cellular networks, where macro, pico, femto base station distributions are modeled via this process \cite{BaccAndGant_2011,DGBA2012}. It is shown that, even though this assumption implies the independent distribution of base stations which may not hold in cellular networks, the model is still accurate and conservative in predicting many important network parameters, while retaining a higher degree of mathematical tractability than standard regular-grid models. 

There have been recent works studying the cooperative schemes employing spatial models. In the context of decentralized networks, in \cite{ARVPG_2013}, the optimal relay activation probability for a network in which source nodes use either direct transmission (DT) or cooperate with a full-duplex relay using DF. In the context of cellular networks, cooperative schemes with spatial models have mainly regarded cooperation between two or more base stations with a wired backhaul. In \cite{BaccGiov_2013Asilomar},
the authors analyze the improvements in coverage probability, achievable through a scheme in which each user can be served by the base stations which is nearest to him, or jointly served by the two nearest ones. On the other hand, in \cite{AndrewsTanb_2013}, the authors use a model based on the Poisson process to analyze the interference distribution in a network employing non-coherent joint-transmission base station cooperation, in which several base stations transmit the same data to a given user, without prior phase mismatch correction. 

\subsection{Main Contributions}
\label{subsec:contributions}
In this paper, we focus on a reference transmitter which attempts to communicate with a destination with the aid of a full-duplex in-band relay, using either the DF or CF schemes. This channel could be interpreted as the downlink of a reference fixed-size cell of a cellular network, in which the base station cooperates with a wireless infrastructure relay. The relay is therefore not connected to the wired backhaul of the network, as in previous works, and receives the message wirelessly from the base station. We consider that the co-channel interference experienced by the relay and the destination of the message comes from nodes which are distributed as Poisson process and that signals are subject to path loss attenuation and Rayleigh fading. In this analysis, we focus on the interference and the signal-to-interference ratio (SIR), neglecting the presence of background noise, since in many scenarios this is the main limit to performance \cite{BaccAndGant_2011}.  For DF, we derive the expression of the exact OP of the link in terms of the joint Laplace transform of the interferences, which unfortunately, can only be evaluated numerically. In addition, we perform an asymptotic analysis of the OP as the density of interferers tends to zero (SIR $\rightarrow \infty$), and show that it is minimized when the correlation coefficient between the source and relay's symbols is zero.
 We then derive closed-form upper bounds to the OP which are tight for small OPs, typical in wireless system designs \cite{BPS_1998}. In the case of DF, the requirement that the relay fully decode the messages of the source prior to forwarding them, implies that the rate that the link can achieve does not depend on the spatial correlation of the interference signals at the relay and the destination. This spatial correlation is a consequence of the spatial distribution of interfering nodes which interfere with the relay and the destination simultaneously. In CF, however, the relay compresses the messages received from the source without decoding them and hence, the achievable rate depends on the spatial correlation induced by the point process. For this reason, a direct analysis of the CF protocol is infeasible. In this work, we show that the achievable rate of CF with correlated interference is at most one bit worse than with uncorrelated interference and, using this fact, we derive an upper bound on the OP of CF. Finally, we also compare the performance of the analyzed protocols with a simple half-duplex DF protocol and with DT. To do this, a lower bound on the OP for a half-duplex DF protocol is introduced.
 
The rest of the paper is organized as follows: in Section \ref{sec:model}, 
we present the mathematical model of the network, a description of the DF and CF schemes, and their achievable rates.
The OP is also shown to be an upper bound on the asymptotic error probability of the link. In Section \ref{sec:nonhomogeneous} the OP analysis is carried out for both schemes. Finally, numerical results and conclusions can be found in Sections \ref{sec:numerical} and \ref{sec:final}, respectively, while proofs are relegated to the appendices.

\subsection*{Notation} We denote as $\mathbb{R}$, $\mathbb{C}$ and $\mathbb{R}^2$, the real numbers, complex numbers and the real plane respectively.
The Euclidean norm is denoted as $\|\cdot\|$. $(\cdot)^{*}$ denotes complex conjugation and $\Re(\cdot)$ the real part of complex number. $\prob_X\left\{\cdot\right\}$ and $\Ex_X\left[\cdot\right]$ denote probability and expectation with respect to the random variable $X$. $\mathcal{A}^{c}$ and $\mathds{1}\{\mathcal{A}\}$ denote the complement and the indicator function of the set $\mathcal{A}$, and $I(X,Y)$ denotes the mutual information between the random variables $X$ and $Y$. We use the big O notation: $f(x)=\Onot\left(g(x)\right)$ as $x\rightarrow x_0$ if there exists $M>0$ such that $|f(x)|\leq M|g(x)|$ in some neighborhood of $x_0$.
\section{General considerations and network model}
\label{sec:model}
\subsection{Spatial Model and Preliminaries}
\label{subsec:math_descrip}
We consider a single-antenna source node located at the origin $o \in \R^2$ which attempts to communicate with a destination located at $d = (D,0)$ with the aid of a relaying node located at $r$, working in a full-duplex fashion in the same frequency band. We model the interfering nodes as an independently marked homogeneous Poisson point process \cite{stochastic_geometry2009}:
\begin{equation}
\tilde{\Phi} = \left\{ (x_i,h_{x_ir},h_{x_id})\right\}, \label{eq:PPP}
\end{equation}
with the following characteristics:
\begin{itemize}
\item The set of transmitters constitutes an homogeneous Poisson process $\Phi = \left\{ x_i \right\}$ of intensity $\lambda$.    
\item All users transmit with constant unit power. We assume that transmissions are affected by path loss and i.i.d. narrow-band block-fading, that is, the power received at $y$ by a transmitter at $x$ is $|h_{xy}|^2 l(x,y)$ where:
\begin{itemize}
\item $l(x,y)$ is a spherically symmetric path loss between $x$ and $y$. For numerical results we shall work with the usual unbounded or \emph{simplified} path loss function: $l(x,y) = ||x-y||^{-\alpha}$ with $\alpha >2$.
\item $|h_{xy}|^2$ is the power fading coefficient associated with the channel between points $x$ and $y$. We consider Rayleigh fading, i.e. the power fading coefficients are independent identically distributed exponential random variables with unit mean.
This is equivalent to saying that $h_{xy}$ are complex, circular, zero-mean Gaussian random variables.
\end{itemize}
\item The marks $h_{x_ir}$ and $h_{x_id}$ model the fading coefficient between each transmitting node in the network and the nodes relay and destination corresponding to the transmitter located at the origin, respectively. In addition we include another fading coefficient $h_{rd}$ with the same distribution as $h_{x_ir}$ and $h_{x_id}$, independent of $\tilde{\Phi}$, which models the fading between the relay and destination corresponding to the transmitter at the origin. We denote by $l_{sd}$, $l_{sr}$ and $l_{rd}$ the source-destination, source-relay and relay-destination path losses respectively.
\end{itemize}
Conditioning on the marked point process, the signals received at the relay and destination, associated with the source node at the origin, can be written as:
\begin{equation*}
Y_r  = h_{sr}\sqrt{l_{sr}} X_s + \underbrace{\sum_{i:x_i\in\Phi} h_{x_ir}{l(x_i,r)}^{\frac{1}{2}}X_{x_i}}_{\tilde{Z}_r}
\end{equation*}
\begin{equation*}
Y_d= h_{sd}\sqrt{l_{sd}}X_s+h_{rd}\sqrt{l_{rd}}X_r
+ \underbrace{\sum_{i:x_i\in\Phi}\!h_{x_id}l(x_i,d)^{\frac{1}{2}}X_{x_i}}_{\tilde{Z}_d},
\end{equation*}
where, for shortness, we have dropped the dependence of the signals on the messages to be transmitted and the discrete time indices for the block codewords. We have denoted with $(X_s,X_r)$ the symbols transmitted by the source and the relay and  $\{X_{x_i}\}$  the corresponding signals for the other transmitters in the network. If Gaussian signaling is used, that is, the $\{X_{x_i}\}$ are generated as complex, circular, unit-variance, zero-mean independent Gaussian random variables, and interference is treated as noise for decoding, then for each realization of $\tilde{\Phi}$ the aggregate interferences $\tilde{Z}_r$ and $\tilde{Z}_d$, are zero-mean complex circular Gaussian variables whose conditional variances are:
\begin{gather}
I_r  \overset{\vartriangle}{=} \Ex \left[|\tilde{Z}_r|^2 \vert _{\tilde{\Phi}}\right] = \sum_{i:x_i\in\Phi} |h_{x_ir}|^2l(x_i,r) ,
\label{eq:power_inter_relay} \\
I_d \overset{\vartriangle}{=} \Ex \left[|\tilde{Z}_d|^2\vert _{\tilde{\Phi}}\right] = \sum_{i:x_i\in\Phi} |h_{x_id}|^2 l(x_i,d).
\label{eq:power_inter_dest}
\end{gather}
In addition, these signals are spatially correlated by the point process, and their correlation coefficient is:
\begin{equation}
\rho_N\vert_{\tilde{\Phi}} = \frac{\Ex[\tilde{Z}_r \tilde{Z}_d^* \vert _{\tilde{\Phi}}]}{\sqrt{I_r I_d}}. \label{eq:rhon1}
\end{equation}
As we mentioned previously, depending on the protocol used, this correlation may have an impact on the achievable rate and it may be necessary to take it into account in the analysis. 
We can define the joint Laplace transform for these interference power random variables $I_r$ and $I_d$ as:
\begin{equation}
\mathcal{L}_{I_d,I_r}\left( \omega_1, \omega_2\right) := \Ex_{\tilde{\Phi}} \left[ e^{-(\omega_1 I_d + \omega_2 I_r)} \right],\ \omega_1,\omega_2\in\mathbb{C},
\label{eq:lapdef}
\end{equation}
with $\Re\left\{\omega_1\right\}, \Re\left\{\omega_1\right\}>0$. The general expression for the Laplace transform of interference random variables of an independently marked Poisson process can be found in \cite[chap. 2]{BB2010}. In our case, given (\ref{eq:power_inter_relay}) and (\ref{eq:power_inter_dest}) we have:
\begin{align}
\mathcal{L}_{I_d,I_r}(\omega_1,\omega_2) 
&=  \exp\left\{ -\lambda \int_{\R^2}  \Ex\left[1- \hspace{-2pt}e^{ -\omega_1 g(x,d,h_1) -\omega_2  g(x,r,h_2)}\right] dx\right\}, \\
&=\exp\left\{-\lambda \hspace{-1mm}\int_{\R^2} \hspace{-1mm} \left[ 1- \frac{1}{(1+\omega_1l(x,d))(1+\omega_2l(x,r))} \right] dx  \right\},\label{eq:Lapgen1} \hspace{-3mm}
\end{align}
where $g(x,y,h)=l(x,y) |h|^2$ is the function appearing inside the interferences. In the second step we computed the expectation over $\{|h_1|^2, |h_2|^2\}$, which are unit mean independent exponential random variables representing the fading coefficients inside the interference random variables. 
\begin{Lemma} \label{lem:LapTransSimple}
For the simplified path loss function the Laplace transform writes as:
\begin{equation}
\mathcal{L}_{I_d,I_r}(\omega_1, \omega_2) = e^{-\lambda \left(C (\omega_1^{2/\alpha}+ \omega_2^{2/\alpha}) + f(\omega_1, \omega_2) \right)}, \label{eq:Lap2D}
\end{equation}
where:
\begin{equation}
C = \frac{2\pi\Gamma\left(\frac{2}{\alpha}\right) \Gamma\left(1-\frac{2}{\alpha}\right)}{\alpha} \label{eq:C}
\end{equation}
\begin{equation}
f(\omega_1 , \omega_2) = \int_{\R^2} \frac{\omega_1 \omega_2}{(\omega_1+ \|x-d||^\alpha) (\omega_2+ ||x-r||^\alpha)} dx \label{eq:integral_hard}
\end{equation}
and $\Gamma(z) = \int_0^\infty t^{z-1} e^{-t} dt$ is the Gamma
function. 
\begin{IEEEproof}
See appendix \ref{proof:LapTransSimple}.
\end{IEEEproof}
 $f(\omega_1,\omega_2)$ in (\ref{eq:integral_hard}) accounts for the statistical dependence between the interferences at two different locations, which is inherited from the spatial correlation of the time signals $Y_r$ and $Y_d$. If, in fact, they were independent,  the joint Laplace transform would be the product of their individual transforms and this cross-term would not appear. Unfortunately this term does not have a closed form and is very difficult to bound tightly in a general setup for all $(\omega_1,\omega_2)$. Taking $\omega_1=0$ or $\omega_2=0$, the separate Laplace transforms of the interference powers, which have closed form expressions for the simplified path loss function, can be obtained.
\end{Lemma}
\vspace{-5mm}
\subsection{Problem Statement and Bounds on the Asymptotic Error Probability}
\label{subsec:error_asymp}

Our goal is to study the asymptotic error probability performance of the relay channel formed by the source at the origin together with its destination and the relay.  We assume there is no channel state information (CSI) available at the source, that the relay only has CSI of the source-relay channel, and the destination of both the source-destination and relay-destination channels. 
A message $W$ is chosen at random by the source, and a transmission takes place using a single-relay code, defined as follows:
\begin{definition}[single-relay code]
\label{def:code}
A  single-relay code $\mathcal{C}_n(n,M_n)$ of rate $R$ for a set of   messages $\{1,\dots,M_n\}$ consists of:
\begin{itemize}
\item A set of random and independent complex Gaussian  codewords $X_s^n(w_i)$, $w_i \in \{1,\dots,M_n\}$, each according to $n$ i.i.d. draws of a unit-variance Gaussian variable.
\item A decoder mapping $\hat{W} :\mathbb{C}^{n} \longmapsto \{1,\dots,M_n\}\cup \{\mathcal{E}\}$.
\item A sequence of relay mappings $f_{t} :\mathbb{C}^{t-1} \longmapsto \mathbb{C} $ constrained to produce i.i.d. complex Gaussian random variables of unit variance, for $t=\{1,\dots,n\}$.
\end{itemize}
\end{definition}

The smallest asymptotic average (over all random parameters) probability of error of the source-destination pair at the origin is given by:
\begin{equation}
\bar{P}_{e}(R,\lambda) \equiv \\ \inf_{\mathcal{C}_{n}} \Big\{\limsup_{n\rightarrow \infty}\mathbb{P}_{\Theta,\lambda}^{(n)}(W\neq \hat{W} | \mathcal{C}_{n}) \, \Big| \liminf_{n\rightarrow \infty} \frac{1}{n}\log M_{n}\geq R \Big\},
\end{equation}	
where $\Theta$ condenses all the randomness in the model:
\begin{equation}
\Theta=\left\{\tilde{\Phi},h_{sr},h_{rd},h_{sd},r\right\}.
\label{eq:Theta}
\end{equation}
Since the source is unaware of the instantaneous interference, path loss attenuation and fading coefficients involved, the error probability cannot be made arbitrary small with the code-length. For any code $\Cn$ according to Def. \ref{def:code}, the error probability can be bounded as \cite{BPS_1998}:
\begin{align}
\prob(W\neq\hat{W}|\Cn) &= \prob(W\neq\hat{W}|\Cn, \mathcal{O}(R)) \prob(\mathcal{O}(R)|\Cn) +
\prob(W\neq\hat{W}|\Cn, \mathcal{O}^c(R)) \prob(\mathcal{O}^c(R)|\Cn) \nonumber\\
&\leq \prob(\mathcal{O}(R)|\Cn) + \prob(W\neq\hat{W}|\Cn, \mathcal{O}^c(R)),
\end{align} 
where $\mathcal{O}(R) \in \sigma(\Theta)$ denotes an outage event, $\sigma(\Theta)$ being the $\sigma$-algebra generated by $\Theta$. Therefore, the asymptotic error probability can be upper bounded by any code $\Cn$, as follows:
\begin{equation}
\bar{P}_{e}(R,\lambda) \leq \inf_{\mathcal{O}(R)\in\sigma(\Theta)} \left[ \prob_{\Theta,\lambda} \left\{\mathcal{O}(R)\right\}  +  \limsup_{n\rightarrow \infty} \prob^{(n)}_{\Theta,\lambda} \left\{ W\neq \hat{W} |\mathcal{C}_n, \mathcal{O}^c(R)\right\} \right].
\label{eq:General-outage}
\end{equation}
If, for a given code $\mathcal{C}_n$, the event $\mathcal{O}^c(R)$ is chosen to contain the points in $\Theta$ such that $R$ is achievable through $\Cn$, the second term on the right-hand side of  \eqref{eq:General-outage} can be made arbitrary small. That is,  for any $\epsilon>0$:
\begin{equation}
\limsup_{n\rightarrow \infty} \prob^{(n)}_{\Theta,\lambda} \left\{ W\neq \hat{W} |\mathcal{C}_n,\mathcal{O}^{c}(R)\right\}\leq \epsilon.
\label{eq:given-outage}
\end{equation}
In this way, given a rate $R$, the asymptotic error probability $\bar{P}_{e}(R)$ is dominated by the OP $\prob_{\Theta,\lambda} \left\{\mathcal{O}(R)\right\} $ of the corresponding achievable rate. The OP is a useful performance metric which was extensively employed to characterize performance in a Poisson field of interferers, jointly with the associated metric of \emph{transmission capacity} \cite{haenggi_stochastic_2009,weber_overview_2010}. 

In what follows, for shortness we write $\prob_{\Theta,\lambda} \equiv \prob$. We  shall also consider the scaling behavior of the error probability with the density of interferers. We have the following definition:
\begin{definition}[small node-density regime]
The following metric $\kappa(R)$ characterizes the asymptotic error probability $\bar{P}_e(R,\lambda)$ as $\lambda\rightarrow 0$, that is, in the high-SIR\cite{HGG2011,Ganti11tit2011}, or small node-density regime:
\begin{equation}
\kappa(R)\equiv\lim_{\lambda\rightarrow 0}\frac{\bar{P}_{e}(R,\lambda)}{\lambda}  \leq \lim_{\lambda\rightarrow 0}\frac{\prob\left\{\mathcal{O}(R,\lambda)\right\}}{\lambda}.
\label{eq:SNDR}
\end{equation}
\label{def:kappa}
\end{definition}
This parameter indicates the behavior of the error probability as the density of interferers tends to zero (SIR $\rightarrow \infty$),  in which case we have, $\bar{P}_{e}(R,\lambda) \approx  \kappa(R) \lambda$. This is a good approximation in the typical, small error probability operating regime of wireless networks \cite{BPS_1998}. Since the best error probability, and hence, $\kappa(R)$, cannot be found in closed form, we can use the upper bound on the right side of (\ref{eq:SNDR}), which is the  \emph{spatial contention} parameter introduced in \cite{HGG2011}. This parameter represents the slope of the OP as $\lambda \rightarrow 0$, and is an upper bound to the best slope attainable, given by $\kappa(R)$.
\vspace{-5mm}
\subsection{Achievable Bounds on the Asymptotic Error Probability}
Cover and El Gamal introduced the main coding strategies for the relay channel in their seminal paper \cite{cover_capacity_1979}: decode-and-forward and compress-and-forward. There exists also a third strategy, amplify-and-forward \cite{laneman_cooperative_2004} which we will not consider.
\subsubsection{Decode-and-Forward}
\label{subsec:DF}

in this protocol, the relay decodes the messages sent by the source, re-encodes them and forwards them to the destination, which employs the transmissions of both users to decode the message. In the special case where the memoryless relay channel is \emph{physically degraded} the achievable rate using DF is in fact the capacity. In the general case there is not a unique scheme maximizing the rate for all channel parameters. Since the standard version of DF requires the relay to fully decode the message of the source, this strategy will work best when this channel is good enough with respect to the source-destination channel so that a bottleneck is not introduced. In a scenario in which the spatial attenuation of signals is considered, this will happen when the relay is on average closer to the source than to the destination. Other variants of DF such as partial decode-and-forward \cite{el_gamal_bounds_2006} partially overcome this requirement, but they require an optimization of the code at the encoder, which cannot be done unless CSI is available at the source.

In order to bound the average asymptotic error probability with $\prob \left\{\mathcal{O}(R)\right\} $ as discussed in the previous section, we need to define the outage events associated with DF. There exist several coding schemes which achieve the same DF rate, all based on block-Markov  coding \cite{kramer_cooperative_2005}. In this work we consider the outage events associated to DF with \emph{regular encoding} and \emph{sliding-window decoding} at the destination or \emph{regular encoding} and \emph{backward decoding}. In \cite{cover_capacity_1979}, DF is defined using \emph{irregular encoding}, \emph{random binning} and \emph{successive decoding} at the destination but this strategy has additional error events so it will not be considered. Using  regular encoding and sliding-window decoding \cite{kramer_cooperative_2005} with Gaussian signaling, the $n$-length random codewords at each source and its associated relay are:
\begin{gather}
X_s^n(w_{i-1},w_i)=\sqrt{(1-|\rho|^2)}\tilde{X}_1^n(w_i)+\rho\tilde{X}_2^n(w_{i-1}), \\ X_r^n(w_{i-1})=\tilde{X}_2^n(w_{i-1}),\label{eq:Gaussian_signal_SR}
\end{gather}
for messages $w_i\in\{1,\dots, 2^{nR}\}$ with $w_0=w_{B+1}=1$ and each block $i=\{1,\dots,B\}$. $\tilde{X}_1$ and $\tilde{X}_2$ are independent complex, circular Gaussian random variables with unit variance and $\rho$ is the correlation coefficient between source and relay signals $X_s$ and $X_r$. This protocol is oblivious to the presence of the relay, that is, if the relay does not decode or choses not to transmit, it does not degrade the performance of the protocol with respect to DT \cite{katz_cooperative_2009,BehbPiant2012}. The destination needs to know if the relay will transmit, which is cost-free as the block-length grows. If we define the event that the relay is able to decode source's message:
\begin{equation}
\adf(R,\rho)=\left\{\mathcal{C}\left(|h_{sr}|^2 l_{sr}\left(1-|\rho|^2)/I_r\right)\right)<R \right\}, \label{eq:adfev}
\end{equation}
with $\mathcal{C}(u) = \log_2(1+u)$, then for each realization of $\tilde{\Phi}$ any attempted rate $R$ that satisfies:
\begin{equation}
R \leq \mathds{1}\{\adf(R,\rho)\} R_M(\rho) + \mathds{1}\{\adf^c(R,\rho)\} R_{DT},  \label{eq:achievRDF}
\end{equation}
is achievable, where:
\begin{align}
R_{DT} &= \mathcal{C}(|h_{sd}|^2 l_{sd}/I_d), \label{eq:RDT}\\
R_{M}(\rho) &= \mathcal{C} \!\left(\!\frac{|h_{sd}|^2l_{sd}+|h_{rd}|^2l_{rd}+ 2\sqrt{l_{sd}l_{rd}}\Re\left(\rho
h_{sd}h_{rd}^{*}\right)}{I_d}\right)\!, \nonumber
\end{align}
are the rates of a DT from the source to the destination, and a joint transmission from the source and the relay to the destination, respectively. When $\rho = 0$, the rates of this protocol are the same as those of block-Markov multiplexed coding \cite{HostMad2004,HostMad2006}. Introducing the following outage events:
\begin{IEEEeqnarray}{rCl}
\bdf(R,\rho)&=& \left\{ R_M(\rho) < R \right\}, \\ 
\adt(R) &=& \left\{ R_{DT} < R\right\}, \label{eq:adt}
\end{IEEEeqnarray}
the outage event for this protocol is, from (\ref{eq:achievRDF}):
\begin{equation*}
\mathcal{O}_{DF}(R,\rho)\!=\![\adf^c(R,\rho) \cap \bdf(R,\rho)]  \cup [\adf(R,\rho) \cap \adt],
\end{equation*}
\hspace{-1.5mm}for which condition \eqref{eq:given-outage} holds true. The event $\bdf(R,\rho)$ means that the destination is in outage while source and relay cooperate. The error probability is bounded by:
\begin{IEEEeqnarray}{rcl}
\bar{P}_{e}(R,\lambda) &\leq& \inf_{\rho\in\mathbb{C}, |\rho|\leq 1}\poutdf(R,\rho), \label{eq:evodf}
\end{IEEEeqnarray}
where $\poutdf(R,\rho) = \prob \left\{ \mathcal{O}_{DF}(R,\rho) \right\}$.
Notice that the imposition of full decoding at the relay implies that the achievable rate does not depend on the spatial correlation of the aggregate interference signals at the relay and destination, given by (\ref{eq:rhon1}).

\subsubsection{Compress-and-Forward} In this scheme, the relay compresses the received signal without decoding the message and forwards this compressed description to the destination. There are several coding schemes which, for a given probability mass function $p_{X_s} p_{X_r} p_{\hat{Y}_r|X_r,Y_r}$, achieve the CF rate\cite{el_gamal_bounds_2006}:
\begin{equation}
R_{CF} = \min \{ I(X_s,X_r;Y_d) - I(Y_r,\hat{Y}_r|X_s,X_r,Y_d),  I(X_s; \hat{Y}_r,Y_d | X_r)\}.\label{eq:rateCf1}
\end{equation}
$\hat{Y}_r$ represents the compressed representation of the symbols received by the relay $Y_r$. As the relay is not compelled to decode the source message, there is not bottleneck in the information flow through the relay as in DF. When the relay is close to the destination CF will compress $Y_r$ and transmit this description to the destination with little effort and CF will typically outperform DF and DT.
In the Gaussian relay channel it is customary \cite{kramer_cooperative_2005} to choose $\hat{Y}_r = Y_r + Z_c$ and $X_s$, $X_r$ and $Z_c$ independent complex, circular Gaussian random variables with unit variance for the first two and variance $n_c$ for the third one. Thus, the following rate can be achieved  from (\ref{eq:rateCf1}):
\begin{equation}
R_{CF}(\rho_N,n_c) = \min\{ R_1(\rho_N,n_c),R_2(\rho_N,n_c)\}, \label{eq:rateCf2}
\end{equation} 
\begin{equation*}
\hspace{-5.5mm}R_1(\rho_N,n_c)\hspace{-1mm} = \Cfunc \hspace{-.5mm}\left(\hspace{-.5mm}\frac{|\hsd|^2 l_{sd} + |\hrd|^2l_{rd}}{I_d}\hspace{-.5mm}\right)\hspace{-.5mm} -  \Cfunc \hspace{-.5mm}\left(\hspace{-1mm}\frac{I_r}{n_c} (1-|\rho_N|^2)\hspace{-1mm} \right)\hspace{-4mm} \label{eq:R1_r}
\end{equation*}
\begin{equation}
\hspace{-4mm} R_2(\rho_N,n_c) \hspace{-1mm} = \Cfunc\left(\frac{|\hsd|^2l_{sd}}{I_d} + \frac{\frac{|\hsd|^2l_{sd}}{I_d} |\rho_N|^2 +\frac{|\hsr|^2l_{sr}}{I_r}-2 \Re\left\{ \rho_N \frac{\hsd \sqrt{l_{sd}}}{\sqrt{I_r}}\frac{\hsr^*\sqrt{l_{sr}}}{\sqrt{I_d}}
\right\}}{1 + \frac{n_c}{I_r} - |\rho_N|^2} \right)\hspace{-1mm}.\hspace{-1mm} \label{eq:R2_r}
\end{equation}
We have made explicit the dependence of the achievable rate with the spatial noise correlation coefficient $\rho_N$ given by (\ref{eq:rhon1}) to mark a distinction with DF in which this correlation does not affect the rate. Notice that the rate is also dependent on the compression variance $n_c$ of choice. In general, whenever:
\begin{equation}
I(X_r;Y_d) \geq I(Y_r; \hat{Y}_r | X_r, Y_d), \label{eq:condNc}
\end{equation}
then the rate $R_{CF}$ is the second term in (\ref{eq:rateCf2}) \cite{el_gamal_bounds_2006}, that is:
\begin{equation}
R_{CF} (\rho_N) = I(X_s;\hat{Y}_r, Y_d | X_r).
\end{equation}
For the Gaussian relay channel, after choosing $X_s$, $X_r$ and $Z_c$ as indicated above (\ref{eq:rateCf2}), condition (\ref{eq:condNc}) is \cite{cover_capacity_1979}:
\begin{equation}
n_c \geq \frac{I_r I_d}{|\hrd|^2l_{rd}}  \left(\frac{|\hsd|^2 l_{sd}}{I_d}+ \frac{|\hsr|^2l_{sr}}{I_r} -  2\Re \left\{\rho_N \frac{\hsd \hsr^* \sqrt{l_{sd}l_{sr}}}{\sqrt{I_r I_d}} \right\}+1\right). \label{eq:Nc1}
\end{equation}
This means that is we define the event $\bcf(n_c,\rho_N) = \{ \text{(\ref{eq:Nc1}) is not met}\}$, then any rate $R$ that satisfies:
\begin{equation}
R < \mathds{1}\{\bcf^c(n_c,\rho_N,n_c)\} R_2(\rho_N,n_c),
\end{equation}
is achievable \cite{cover_capacity_1979}. Condition  (\ref{eq:Nc1}) implies that the relay-destination channel can sustain the rate to transmit the compressed version of what the relay receives from the source.
If full CSI is available at the relay, we can choose the compression variance $n_c$ to achieve equality in (\ref{eq:Nc1}). Otherwise, the value of $n_c$ has to be fixed a priori and an outage event will take place when the realization of the network does not allow (\ref{eq:Nc1}) to be fulfilled. Therefore, we can define the outage event $\mathcal{O}_{CF}(R,\rho_N,n_c)=\left\{\acf(R,\rho_N)\cup \bcf(n_c,\rho_N)\right\}$, with:
\begin{align}
\acf(R,\rho_N,n_c) &= \left\{ R_2(\rho_N,n_c) < R \right\}.
\end{align}
Notice that CF cannot perform worse than DT, because by taking an arbitrarily large value of $n_c$ we guarantee that (\ref{eq:Nc1}) will be met, and in that case, inspecting (\ref{eq:R2_r}) we check that $R_2(\rho_N,n_c)$ will be very arbitrarily close to the rate of DT.
\subsubsection{Half-duplex DF}
In order to compare the performance of the full-duplex DF protocol we introduced before, we also consider a half-duplex DF strategy, known as \emph{sequential} DF \cite{katz_cooperative_2009}. In this scheme, the transmission is split in two phases. In the first one, occupying a fraction $0 \leq \varepsilon < 1$ of the block, the source transmits its message to the destination while the relay listens and attempts to decode the message. If the relay is able to decode the message during the first phase, it employs the remaining $(1-\varepsilon)$ fraction of the block to transmit, acting as a secondary antenna. This scheme is also oblivious, so if the relay does not decode the message, it does not degrade the performance with respect to DT. The event that the relay does not decode in the first phase is:
\begin{equation}
\asdf(R,\varepsilon) = \{ \varepsilon \ \mathcal{C}(|\hsr|^2l_{sr}/ I_r) < R \}.
\end{equation}
Hence, any rate $R$ which satisfies:
\begin{equation*}
R < \mathds{1}\{\asdf^c(R,\varepsilon)\} R_{SDF}(\varepsilon) + \mathds{1}\{\asdf(R,\varepsilon)\} R_{DT},
\end{equation*}
is achievable, where $R_{DT}$ is given by (\ref{eq:RDT}) and\cite{katz_cooperative_2009}:
\begin{equation*}
R_{SDF}(\varepsilon) \!=\! \varepsilon \ \mathcal{C}\!\left(\!\frac{|\hsd|^2 l_{sd}}{I_d}\! \right) + (1-\varepsilon) \mathcal{C}\! \left(\!\frac{|\hsd|^2l_{sd} + |\hrd|^2 l_{rd}}{I_d}\!\right)\!. \label{eq:Rsdf}
\end{equation*}
The outage event for this protocol is therefore:
\begin{equation}
\mathcal{O}_{SDF} (R,\varepsilon) = [\asdf^c(R,\varepsilon) \cap \bsdf(R,\varepsilon) ] \cup [\asdf(R,\varepsilon) \cap \adt(R)],
\end{equation}
where $\bsdf(R,\varepsilon)= \{ R_{SDF}(\varepsilon)< R\}$. The value of $\varepsilon$ cannot be adjusted for the instantaneous realization of the network, but can be selected a priori, for example, to minimize the OP.
\subsubsection{Direct transmission} We also define the outage event $\mathcal{O}_{DT}(R)=\adt(R)$ given by (\ref{eq:adt}) for the case in which there is no relay and thus the source simply uses DT. The OP for this scheme is known to be \cite{baccelli_aloha_2006}:
\begin{equation}
\poutdt = \prob(\mathcal{O}_{DT}(R)) = 1- \exp\{-\lambda \delta D^2\}, \label{eq:opdt}
\end{equation}
where, using $C$ given by (\ref{eq:C}), we defined:
\begin{equation}
\delta = C(2^R -1)^{2/\alpha}. \label{eq:delta_def}
\end{equation}
\section{Outage behavior}
\label{sec:nonhomogeneous}
In this section we analyze the outage behavior of the relay channel. In the case in which only Gaussian background noise and Rayleigh fading are considered (without interference) very interesting gains have been observed in terms of the OP \cite{laneman_cooperative_2004} \cite{host-madsen_capacity_2005}. In the scenario in which interference comes from a network of interferers, however, we must average over all possible configurations of interfering nodes, considering numerous situations in which communications are severely impaired due to the presence of heavy interference. As we will see, this results in performance gains which are not as large as in the case in which only fading and noise are considered. 
\subsection{Decode-and-forward} \label{subsec:fixed}
We start by considering the DF protocol. In this setup we derive the OP and tight upper bounds under typical network operating conditions. We also determine the  symbol correlation coefficient $\rho$ which minimizes the OP in the small node-density regime. To analyze the OP (\ref{eq:evodf}), it is convenient to rewrite in terms of the success events as:
\begin{equation}
\hspace{-0mm}\poutdf(R,\rho) = \hspace{-1mm}1- \prob\left(\adf^{c}(R,\rho)\cap\bdf^{c}(R,\rho)\right)  - \prob \left( \adf(R,\rho) \cap \adt^c\right).
\label{eq:first_out_fixed_realy}
\end{equation}
It is interesting to mention that the probability of $\bdf$ in (\ref{eq:first_out_fixed_realy}) has two different expressions according to the relay position $r$ and the correlation coefficient $\rho$ of the symbols transmitted by the source and the relay. However, as we shall see, working with only one of them is enough for characterizing the OP behavior. 
\begin{Theorem}[OP of DF] \label{lem:pdffixed} 
 The probabilities involved in the OP of DF (\ref{eq:first_out_fixed_realy}) can be found as follows.
When $||r-d|| \neq D$ or $\rho \neq 0$ we have:
\begin{equation}
\prob\left(\adf^{c}(R,\rho)\cap\bdf^{c}(R,\rho)\right) = 
  \frac{\mu_2 \mathcal{L}_{I_d,I_r} \left(\frac{T}{\mu_2},\frac{T}{\mu_3}\right)- \mu_1 \mathcal{L}_{I_d,I_r} \left(\frac{T}{\mu_1}, \frac{T}{\mu_3} \right)}{\mu_2-\mu_1}, \label{eq:PoutDF2}
\end{equation}
where:
\begin{gather}
\mu_{1} = \frac{1}{2} \left[l_{sd} + l_{rd} - \left( (l_{sd} - l_{rd})^2 + 4 l_{sd} l_{rd} |\rho|^2 \right)^{\frac{1}{2}}\right],  \label{eq:mu1}\\
\mu_{2} = \frac{1}{2}\left[(l_{sd} + l_{rd}) + \left( (l_{sd} - l_{rd})^2 + 4 l_{sd} l_{rd} |\rho|^2 \right)^{\frac{1}{2}}\right], \label{eq:mu2}\\
\mu_3 = l_{sr}\left(1-|\rho|^2\right) \label{eq:mu3},
\end{gather}
and $\mathcal{L}_{I_d,I_r}\left( \omega_1, \omega_2\right)$, the Laplace transform of the interferences, is given by (\ref{eq:lapdef}), and $T = T(R) = 2^R-1$.
In addition, when $||r-d||= D$ and $\rho = 0$, we have $\mu_1 = \mu_2$ and:
\begin{equation}
 \prob\left(\adf^{c}(R,\rho)\cap\bdf^{c}(R,\rho)\right) =\mathcal{L}_{I_d,I_r} \left(\frac{T}{\mu_1},\frac{T}{\mu_3}\right) - \frac{T}{\mu_1} \frac{d \mathcal{L}_{I_d,I_r} (\omega_1,T/\mu_3)}{d\omega_1}\biggl|_{\omega_1 = T/\mu_1}.\label{eq:PoutDF3}
\end{equation}
Finally: 
\begin{equation}
\prob \left( \adf \cap \adt^c\right) = \Lap_{I_d}\left(\frac{T}{l_{sd}}\right) - \Lap_{I_d,I_r} \left(\frac{T}{l_{sd}},\frac{T}{\mu_3} \right). \label{eq:OPDFT2}
\end{equation}
\end{Theorem}
\begin{IEEEproof} 
See Appendix \ref{proof:pdffixed}.
\end{IEEEproof}
Notice that the OP depends only on the absolute value of $\rho$ and not on its phase. This is a consequence of the uniform phase of the Rayleigh fading coefficients \cite{kramer_cooperative_2005,ZCL_2008}.
The fact that the OP has two different expressions comes from the fading distribution that the destination sees from the joint source-relay transmission when $\rho = 0$ is different whether the source and relay are equidistant from the destination or not. However, the expression of the OP when $\rho = 0$ and $||r-d|| = D$, given by (\ref{eq:PoutDF3}), can be obtained by continuously extending the other expression (\ref{eq:PoutDF2}) at these points. Therefore, in what follows we shall focus our interest on (\ref{eq:PoutDF2}) which fully characterizes the OP. 
Unfortunately, the OP cannot be evaluated to a simple expression, mainly because the integral in (\ref{eq:Lapgen1}) does not have a closed form. However, it can be evaluated numerically without difficulty. Still, we can attempt to find the value of $|\rho|$ which minimizes the OP, by studying how this parameter appears in (\ref{eq:PoutDF2}) and (\ref{eq:OPDFT2}). Among other things, $\rho$ allows some control over using DF or DT; this is because as $|\rho| \rightarrow 1$, $\adf$ given by (\ref{eq:adfev}) has a higher probability and hence, DF is used less. In the scenario in which transmissions are limited by independent background noises at the relay and the destination, there exists a value of $|\rho|$, generally non-zero, that minimizes the OP for each value of the system parameters \cite{katz_cooperative_2009,kramer_cooperative_2005}. Although this is true for each realization of $\tilde{\Phi}$, the same may not be true after averaging over the interference. Unfortunately, a general analysis of the optimal $|\rho|$ is not possible, though we can consider the small node-density regime:
\begin{Lemma}[Optimal $\rho$] \label{lem:optrho}
In the small node-density regime, 
\begin{equation*}
\kappa(R) \leq \inf_{0 \leq |\rho| \leq 1} \lim_{\lambda\rightarrow 0}\frac{ \mathcal{O}_{DF}(R,\rho,\lambda)}{\lambda}
= \lim_{\lambda\rightarrow 0}\frac{ \mathcal{O}_{DF}(R,0,\lambda)}{\lambda}.
\end{equation*}
This implies that, at least in the operating condition of the network, the OP when using DF is minimized by taking $\rho = 0$.
\end{Lemma}
\begin{IEEEproof}
See appendix \ref{Ap:rho_optimal}.
\end{IEEEproof}
As pointed out in \cite{kramer_cooperative_2005} (see remark 42) and \cite{host-madsen_capacity_2005}, using $\rho=0$ simplifies the implementation of DF because symbol synchronization between the source and its relay is not required. In what follows we focus in the small node-density regime and the simplified path loss function, and considering Lemma \ref{lem:optrho} we develop bounds on the OP taking $\rho = 0$. If numerical computation of the OP for other values is needed, it is straightforward to use Lemma \ref{lem:LapTransSimple} in Theorem \ref{lem:pdffixed}. Notice that if we take $\rho = 0$, we have $\bdf \subset \adt$ and hence, the protocol is always better than DT. It is straightforward to verify that in this case the OP can be bounded as:
\begin{align}
\poutdf(R,0) &\leq \min\{ \prob(\adf(R,0)) , \prob(\adt \cap \bdf(R,0)^c)\} +\prob(\bdf(R,0)) \nonumber\\
&= \min \{\prob(\adt),\prob(\adt(R,0))+\prob(\bdf(R,0))\}. \label{eq:UBfixedrelay}
\end{align}
This bound will be a good approximation, when the relay is close to source since a close inspection of $\adf$ and $\bdf$ shows that in this setting the event $\bdf$ will be dominant and $\adf$ will have a relatively small probability of occurrence. The following Theorem deals with evaluating the bound (\ref{eq:UBfixedrelay}) and with bounding  $\kappa(R)$ for the small node-density regime:
\begin{Theorem} [OP upper bounds for DF] \label{lem:OPbsingrf} When $||r-d|| \neq D$, considering the simplified path loss function, the OP for $\rho = 0$ can be upper bounded as:
\begin{multline}
\hspace{-3mm} \poutdf(R,0) \leq  \min \left\{1 - e^{-\lambda \delta D^2}, \left(1 - e^{- \lambda\delta||r||^2}\right) +   \right.\\
+ \left. \left(1 - \frac{ D^\alpha e^{-  \lambda\delta ||r-d||^2} - ||r-d||^\alpha e^{-  \lambda\delta D^2}}{ D^\alpha - ||r-d||^\alpha}\right)\right\}, \label{eq:PBDF3} 
\end{multline}
with $\delta$ given by (\ref{eq:delta_def}). In addition, in the small node-density regime, $\kappa(R)$ can be bounded as:
\begin{equation}
\kappa_{DF}(R)\hspace{-.5mm}  \leq \delta \min\left\{D^2, \hspace{-.5mm}
\|r\|^2+ \|r - d\|^2 D^2 \frac{D^{\alpha-2} - \|r - d\|^{\alpha - 2}}{D^\alpha - \|r - d\|^{\alpha}} \right\}\hspace{-.5mm}.\hspace{-3mm}\label{eq:PBDF4}
\end{equation}
\end{Theorem}
Notice again, that the case where $||r-d||= D$ can be treated via continuity arguments as mentioned above. Finally, the following lower bound on the OP of half-duplex DF will be useful to compare it to the OP of full-duplex DF:
\begin{Theorem} \label{lem:psdffixed} The OP of sequential DF can be lower bounded as: 
\begin{multline}
\hspace{-3mm}\prob(\mathcal{O}_{SDF} (R,\varepsilon))
\geq 1- \left[ \Lap_{I_d}\left(\frac{T}{l_{sd}}\right)\! +\! \mathcal{L}_{I_d,I_r} \left(\frac{T}{\tilde{\mu}_1}, \frac{2^{R/\varepsilon}-1}{l_{sr}} \right)  \right.\\
\hspace{-18mm}\left.+ \frac{\tilde{\mu}_2\mathcal{L}_{I_d,I_r} \left(\frac{T}{\tilde{\mu}_2},\frac{2^{R/\varepsilon}-1}{l_{sr}}\right) -\tilde{\mu}_1  \mathcal{L}_{I_d,I_r} \left(\frac{T}{\tilde{\mu}_1},\frac{2^{R/\varepsilon}-1}{l_{sr}} \right) }{\tilde{\mu}_2-\tilde{\mu}_1} \right], \hspace{-1mm}
\end{multline}
with:
\begin{gather}
\tilde{\mu}_{1} = \frac{1}{2} \left[l_{sd} + (1-\varepsilon) l_{rd} - |l_{sd} - (1-\varepsilon) l_{rd}|\right],\\
\tilde{\mu}_{2} = \frac{1}{2} \left[l_{sd} + (1-\varepsilon) l_{rd} + |l_{sd} - (1-\varepsilon) l_{rd}|\right].
\end{gather}
\end{Theorem}
\begin{IEEEproof}
See appendix \ref{proof:psdffixed}.
\end{IEEEproof}
\subsection{Compress-and-forward}
In this section we derive an upper bound for the OP of CF. This analysis is far more involved than that of DF because, as we mentioned earlier, the condition of full decoding that is imposed on the relay for DF results in an achievable rate which does not depend on the spatial correlation of the interference signals at the relay and at the destination. Since in CF the relay generates a sequence which acts as a compressed version of what it receives, without decoding the message from the source, the correlation of the interference signals given by (\ref{eq:rhon1}) does  affect the achievable rate. In \cite{Goldsmith2011}, the authors carry out an analysis of Gaussian relay channels with correlated noises in which the correlation coefficient is fixed and full CSI is available at the relay. This implies that the compression variance $n_c$ can be chosen to achieve equality in (\ref{eq:Nc1}). Under these conditions, the authors compare the performance of CF with correlated and uncorrelated noises and show that negative noise correlation always helps CF, while positive correlation sometimes helps CF. In the setup of this paper, since full CSI is not available at the relay, the compression variance has to be chosen a priori and the additional outage event that (\ref{eq:Nc1}) is not met has to be considered. In addition, it is straightforward to show that under this condition, in which the variance $n_c$ is a fixed constant independent of network parameters, the value of $\rho_N$ which maximizes or minimizes the rate could be located anywhere on the disc $|\rho_N| \leq 1$. Since no closed-form analysis can be carried out considering this random correlation, we resort to a procedure which allows us to bound the effect of the correlation of the interference on the achievable rate.

\begin{Lemma}[Rate gap in CF] \label{lem:GapRates}The achievable rate of CF for any spatial noise correlation $\rho_N$ is at most one bit worse than the rate with uncorrelated noises, that is:
\begin{equation}
R_{CF}(\rho_N,n_c) \geq R_{CF} (0,n_c) - 1.
\end{equation}
\end{Lemma}
\begin{IEEEproof}
See appendix \ref{proof:GapRates}.
\end{IEEEproof}
Using the previous lemma we can work with the OP of CF assuming that the signals are spatially uncorrelated and increase the rate to bound the actual value of the OP, which leads to the following result: 
\begin{Theorem}[OP bound for CF] \label{lem:bouCFOP} 
For an attemped rate $R$ the OP of CF can be upper bounded as:
\begin{align}
\poutcf(R,n_c) 
& \leq \prob(\mathcal{O}_{CF}(R+1,n_c,0)) \nonumber \\
&= \prob(\acf(R+1,0) \cup \bcf(n_c,0)).
\end{align}
In addition we can bound the OP with $\rho_N =0$ as:
\begin{multline}
\prob(\acf(R+1,0)) \leq 1 - e^{-\frac{T n_c}{l_{sr}}} \Lap_{I_r} \left(\frac{T}{l_{sr}}\right) 
- \left[\sum_{n=0}^{N-1} e^{\frac{nn_cT}{Nl_{sr}}} \Lap_{I_d,I_r} \left(\frac{(N-n)T}{N l_{sd}},\frac{nT}{Nl_{sr}}\right) \right. \\ - \left.  e^{\frac{(n+1)n_cT}{Nl_{sr}}} \Lap_{I_d,I_r} \left(\frac{(N-n)T}{N l_{sd}},\frac{(n+1)T}{Nl_{sr}}\right)\right], \label{eq:probacf}
\end{multline}
and:
\begin{equation}
\prob(\bar{\mathcal{A}}_{CF}(R+1,0) \cap \mathcal{B}_{CF}(n_c,0)) \leq 
  1 \hspace{-0.3mm} - \hspace{-0.3mm} \Ex \hspace{-0.4mm}\left[ \mathcal{L}_{I_d,I_r}\hspace{-0.9mm} \left( \hspace{-0.4mm}\frac{(1+T)l_{sr} |\hsr|^2}{T n_c l_{rd}},\frac{(1+T)l_{sd} |\hsd|^2}{T n_c l_{rd}} \right) \hspace{-0.5mm} \right]. \hspace{-3mm}  \label{eq:boulap}
\end{equation}
The expectation is over $\hsr$ and $\hsd$, with $T = T(R+1) = 2^{R+1} - 1$.
For the simplified path loss we can bound (\ref{eq:boulap}) to obtain:
\begin{equation}
\prob(\bar{\mathcal{A}}_{CF}(R+1,0) \cap \mathcal{B}_{CF}(n_c,0)) \leq  1 - \hspace{-0.5mm} \Ex \hspace{-0.5mm}\left[ \mathcal{L}_{I_d} \! \!	\left( \hspace{-1mm}\frac{(1+T) l_{sr} |\hsr|^2}{T n_c l_{rd}} \! \right)\right] 
\hspace{-0.5mm}\Ex \hspace{-0.5mm}\left[ \mathcal{L}_{I_r}\! \!	\left( \hspace{-1mm}\frac{(1+T) l_{sd} |\hsd|^2}{T n_c l_{rd}} \! \right)\right]. \hspace{-3mm} \label{eq:bouacfbcf}
\end{equation}
\end{Theorem}
\begin{IEEEproof}
See appendix \ref{proof:bouCFOP}.
\end{IEEEproof}
The previous bound has the advantage of allowing us to upper bound the OP of CF in a rigorous manner by considering the gap with the worst case of rate that may be achieved when the spatial correlation between the time signals is the worst possible. Since the correlation of the interference plays a role in the achievable rate it 	is possible that the bound is conservative of the performance of CF.

\begin{figure}[!t]
\centering
\includegraphics[width=\figsize \columnwidth,keepaspectratio,trim= 5mm 0mm 10mm 5mm,clip]{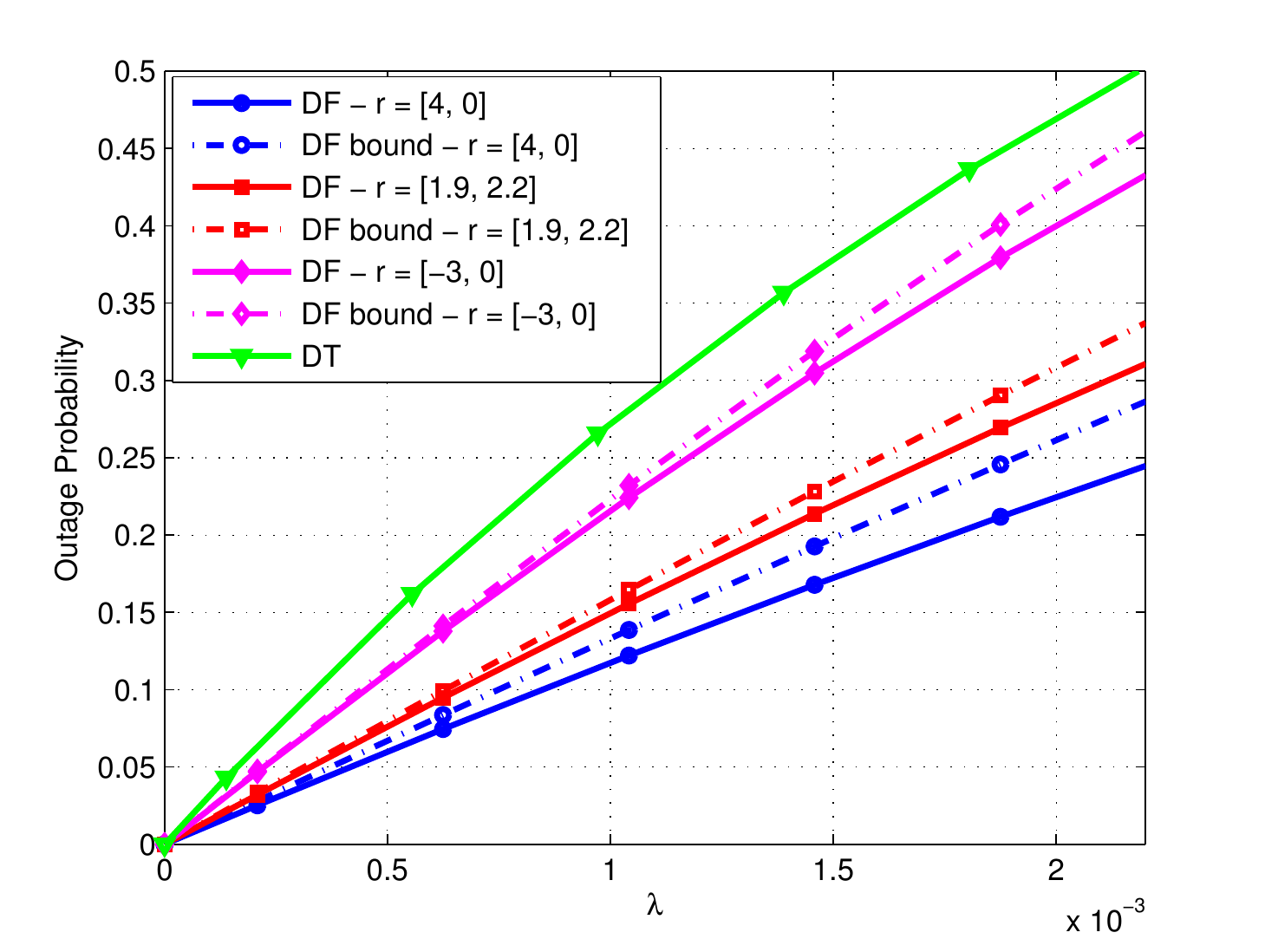} 
\caption{OP of DF with the relay located at $r$ versus DT. Exact expressions for DF come from using Theorem \ref{lem:pdffixed} and Lemma \ref{lem:LapTransSimple}  while upper bounds come from (\ref{eq:PBDF3}). $d = (10 , 0)$, $R = 0.5$ b/use, $\alpha = 4$, $\rho = 0$.}
\label{fig:OPfsr1}
\end{figure}
\section{Numerical Results} \label{sec:numerical}
In this section we present figures to study the  behavior of the derived expressions and to compare the performance of DF and CF with DT and half-duplex DF. In all our simulations we take the destination at $d = (10, \ 0)$ and $\alpha = 4$.

In Fig. \ref{fig:OPfsr1} we can see the comparison of DF versus DT, both through the exact numerical evaluation of the OP using  Theorem \ref{lem:pdffixed} and Lemma \ref{lem:LapTransSimple}, and with the upper bounds given by (\ref{eq:PBDF3}), for different relay positions, taking  $\rho =0$ and $R=0.5$ b/use. We can see that these bounds are accurate when the OP is small,  and the relay is close to the source, as proposed.  In addition, for a fixed source-relay distance the OP increases as the relay grows further away from the destination. This is because the probability of $\bdf(R,\rho)$ increases as this happens. 

\begin{figure}
\centering
\includegraphics[width=\figsize \columnwidth,keepaspectratio,trim= 5mm 0mm 10mm 5mm,clip]{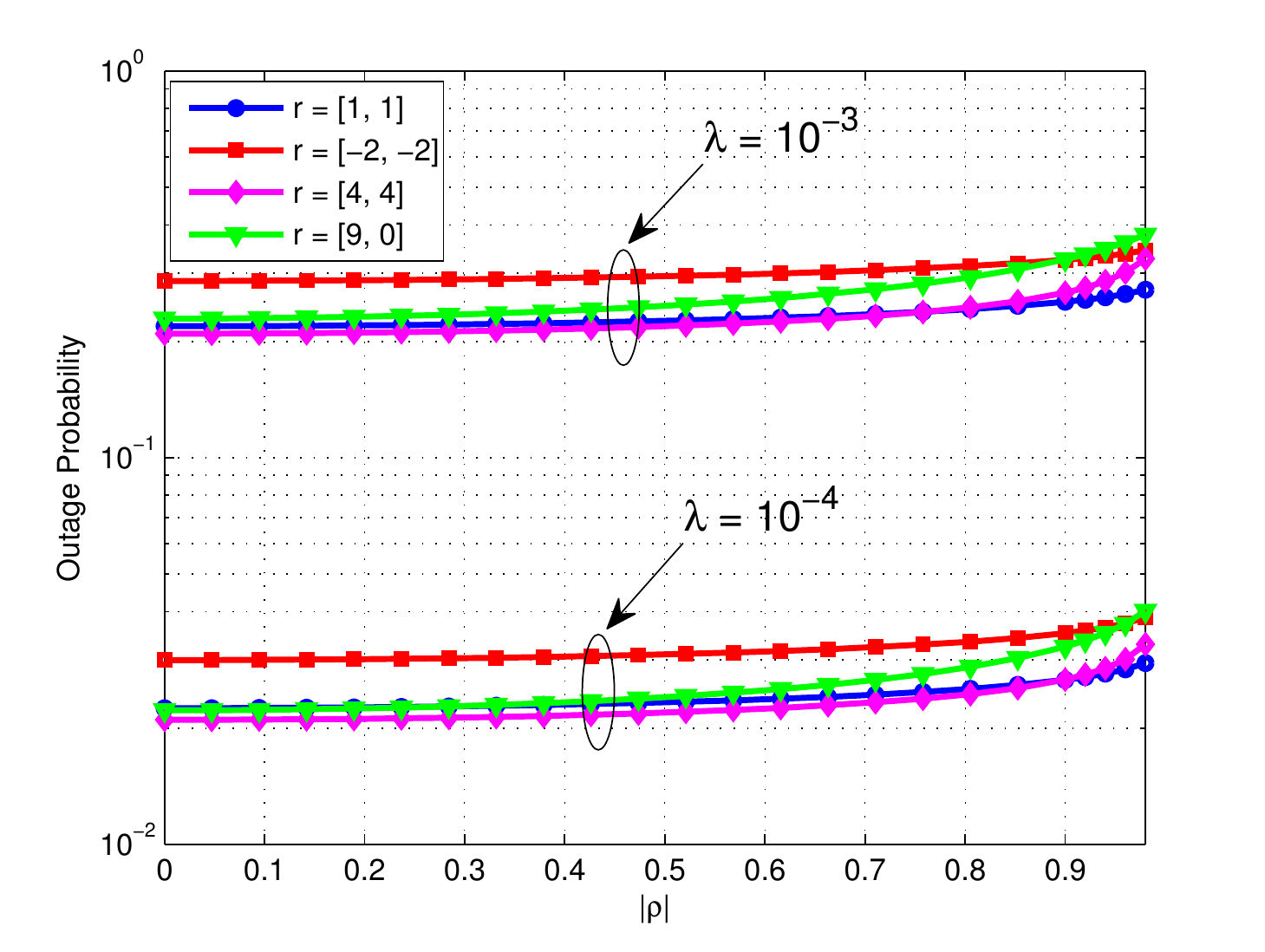} 
\caption{Outage probability as a function of $|\rho|$ for various relay position and for $\lambda=10^{-3}$ and $\lambda=10^{-4}$. $d=(10\ 0)$, $\alpha=4$ and $R=1$ b/use. The OP is found by using Theorem \ref{lem:pdffixed} and Lemma \ref{lem:LapTransSimple}.}
\label{fig:rho_fixed}
\end{figure}

In Fig. \ref{fig:rho_fixed} we can observe how the variation of the true OP of DF given by (\ref{eq:first_out_fixed_realy}) and Theorem \ref{lem:pdffixed}, as a function of the correlation between the symbols of the source and the relay $|\rho|$, for various relay positions, for an attempted rate $R=1$ b/use. Two sets of curves are presented. One for the case of $\lambda=10^{-4}$, in which the OP is small, and the other for $\lambda=10^{-3}$, in which the OP is larger. In both cases we see, as Lemma \ref{lem:optrho} states, $\rho = 0$ is the optimal choice.

\begin{figure}
\centering
\includegraphics[width=\figsize \columnwidth,keepaspectratio,trim= 5mm 0mm 10mm 5mm,clip]{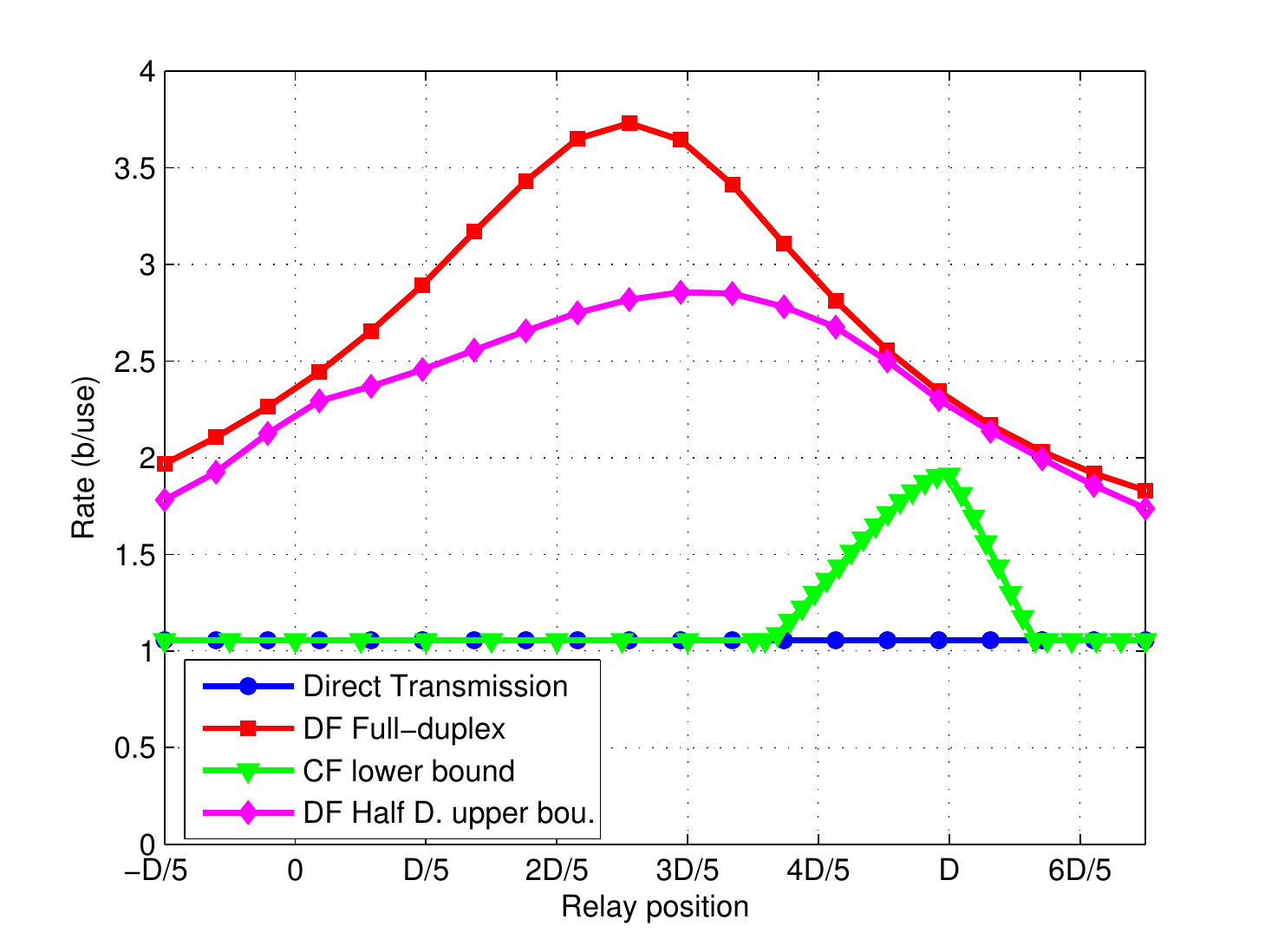} 
\caption{Maximum rate achievable through all the studied protocols when the relay is aligned with the source and destination and an OP smaller that $0.05$ is required. The OP of full-duplex DF comes from Theorem \ref{lem:pdffixed}, half-duplex DF from Theorem \ref{lem:psdffixed}, CF from Theorem \ref{lem:bouCFOP} and DT from (\ref{eq:opdt}). The Laplace transforms are numerical from Lemma \ref{lem:LapTransSimple}. $d = (10 , 0)$, $\alpha = 4$.}
\label{fig:DFVsDT1}
\end{figure}

\begin{figure}
\centering
\includegraphics[width=\figsize \columnwidth,keepaspectratio,trim= 5mm 0mm 10mm 5mm,clip]{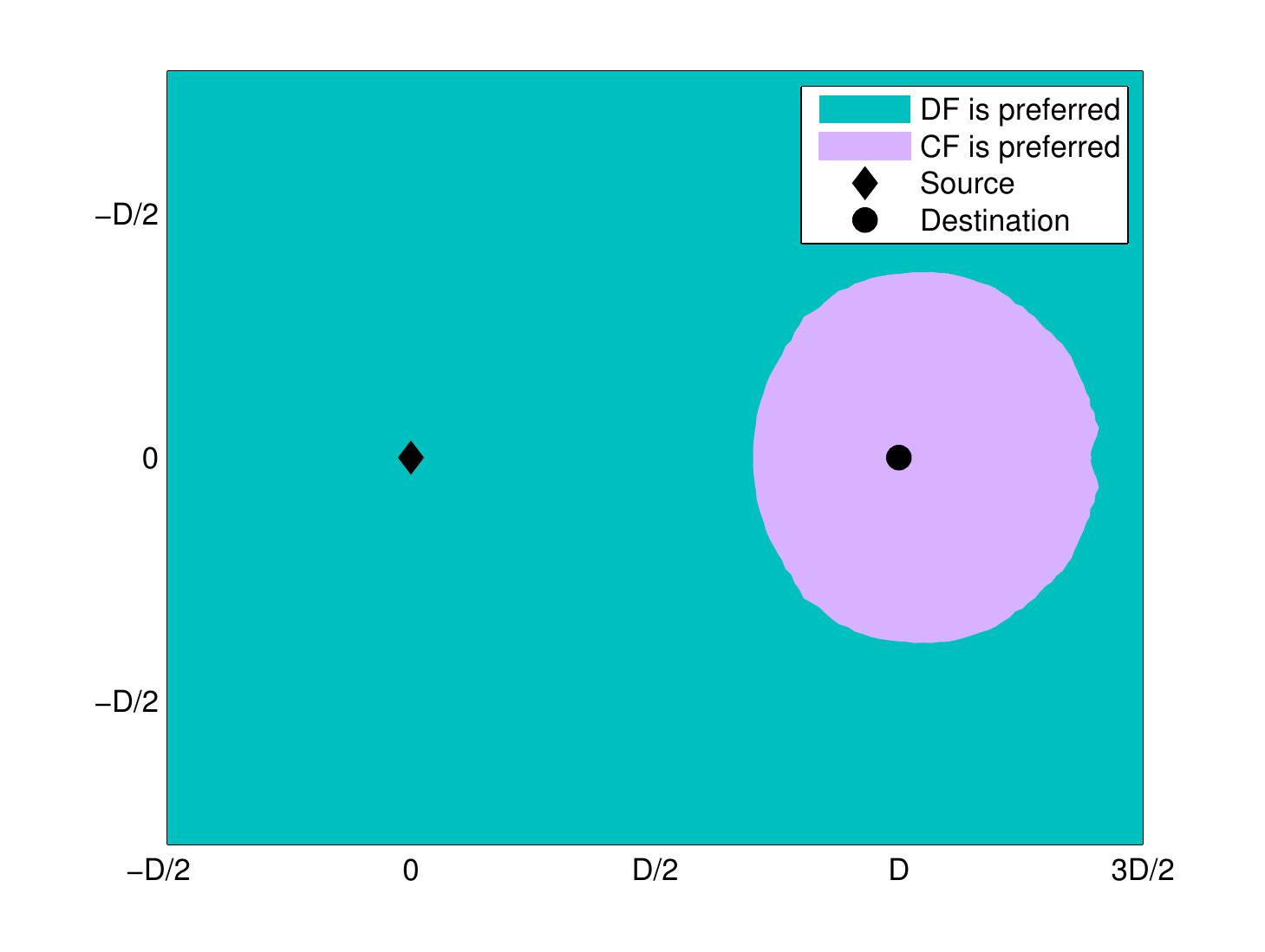} 
\caption{Spatial positions in which each scheme (DF or CF) is preferred. The OP of DF comes from Theorem \ref{lem:pdffixed} with $\rho = 0$, and the performance of CF is estimated by performing a Montecarlo simulation of the point process, optimizing the noise variance $n_c$ and estimating the OP. DT is not shown because it is not better than the other protocols in the plotted region. $\lambda = 0.5 \times 10^{-4}$. $d = (10 , 0)$, $R= 4$b/use, $\alpha = 4$.}
\label{fig:DFVsDT2}
\end{figure}

In Fig. \ref{fig:DFVsDT1} we compare the maximum rate $R$ that can be attempted given a maximum allowed OP of $0.05$, that is, the outage capacity rate for an OP of $0.05$, when the relay is located on the line between the source and destination. We consider full-duplex DF (Theorem \ref{lem:pdffixed}), half-duplex DF (Theorem \ref{lem:psdffixed}), CF (Theorem \ref{lem:bouCFOP}), and DT (Eq. (\ref{eq:opdt})). In the case of half-duplex DF we numerically optimize $\varepsilon$, the fraction of the block in which the relay listens and attempts to decode. The same is done for the compression variance $n_c$ of CF. It can be seen that full-duplex DF outperforms half-duplex DF, specially when the relay is equidistant between the source and destination. The gains however, are not as large near the source or the destination. However, half-duplex DF requires the optimization  of $\varepsilon$ which, as the relay moves away from the source and closer to the destination, takes values on the whole interval $(0,1)$, while for the full-duplex version it suffices to take $\rho = 0$. On the other hand, we see that for CF, the bounds do not predict that CF is better than DF when the relay is near the source, as was observed in other scenarios. This hints that the correlation of the interference may have an important effect on the performance of CF.
In order to explore this, in Fig. \ref{fig:DFVsDT2} we plot the spatial regions in which DF or CF are preferred over the other, for $R= 2$ b/use and $\alpha = 4$. The performance of CF is estimated by performing a Montecarlo simulation of the point process, optimizing the noise variance $n_c$ and estimating the OP. We see that, as expected, CF performs better than DF when the relay is closer to the destination, while DF is better  everywhere else. This shows that, in fact, CF can take advantage of the interference correlation, which impacts its performance significantly, which DF cannot. The optimization of the variance $n_c$, however, is a difficult problem which is not present in full-duplex DF. It is interesting to mention that DT is not shown because it's performance is not close to other protocols in the plotted region.

\section{Summary and Final Remarks}
\label{sec:final}
In this paper we have analyzed the performance, in terms of the OP, of a relay channel employing DF and CF when the interference comes from a network in which nodes are distributed as a Poisson point process. We have derived an expression for the OP of full-duplex DF and upper bounds on the OP which are amenable for analysis and tight when the OP is small. We have also determined the correlation coefficient $\rho$ of the symbols transmitted by the source and the relay which minimizes the OP as the density of interferers vanishes. We showed that the same analysis cannot be carried out for CF, because the achievable rates with this protocol are dependent on the spatial correlation of the interference signals. To avoid this issue, we showed that for any spatial correlation of the point process, the rate achievable is at most one bit worse than the rate when the interferences are uncorrelated. We compared the OP of both full-duplex protocols with half-duplex DF and DT. We have observed that full-duplex DF is the best alternative over the other protocols, except near the destination, where CF is better. However, both CF and half-duplex DF, require the optimization of parameters which depend strongly on the relay's position, while full-duplex DF does not. The disadvantage of the full-duplex protocols comes from the practical aspects involving simultaneous transmission and reception. Hopefully, this analysis may  serve as a starting point for the analysis of more complex network systems and to study the interactions among nodes in large wireless networks, involving different coding and medium-access schemes and network infrastructure.

\appendices
\section{Proof of Lemma \ref{lem:LapTransSimple}} \label{proof:LapTransSimple}
Starting from (\ref{eq:Lapgen1}) we factorize the integrand as:
\begin{equation*}
1 - \frac{1}{( \omega_1 l(x,z_1)+1)( \omega_2 l(x,z_2)+1)} = 
\frac{1}{1+\frac{1}{\omega_1l(x,z_1)}} \\+ \frac{1}{1+\frac{1}{ \omega_2 l(x,z_2)}} 
- \frac{1}{(1+\frac{1}{\omega_1l(x,z_1)})(1+\frac{1}{\omega_2l(x,z_2)})}.
\end{equation*}
The first two terms can be found in closed form:
\begin{equation}
\int_{\R^2} \frac{1}{1+(\omega_1 ||x-r||^{-\alpha})^{-1}}dx = C\omega_1^{2/\alpha}, \label{eq:intbacc1}
\end{equation}
a result which is known from the direct transmission case \cite{baccelli_aloha_2006}. The third term is  (\ref{eq:integral_hard}) after replacing the path loss expression with its expression $l(x,d) = ||x-d||^{-\alpha}$, $l(x,r) = ||x-r||^{-\alpha}$.

\section{Proof of Theorem \ref{lem:pdffixed} } \label{proof:pdffixed}
 In what follows we omit the dependence of the outage event on $(R,\rho)$. We start by (\ref{eq:OPDFT2}), for which we have:
\begin{align}
\prob \left(\adf \cap \adt^c  \right) &= \Ex_{\tilde{\Phi}}\! \left[\prob\left\{|h_{sr}|^2\!<\! \frac{TI_r}{\mu_3},|h_{sd}|^2 \!\geq \!\frac{TI_d}{l_{sd}}\biggl|{\tilde{\Phi}}\right\} \right]\!,\nonumber \\
&= \Ex_{\tilde{\Phi}} \left[ \left(1-e^{-\frac{TI_r}{\mu_3}}\right)e^{-\frac{T I_d}{l_{sd}}}\right], \label{eq:PoutDF1}
\end{align}
where we considered that the power fading coefficients are independent exponential random variables. Applying the definition of the Laplace transform (\ref{eq:lapdef}) we have (\ref{eq:OPDFT2}). For (\ref{eq:PoutDF2}) we define:
$Z:= |{h}_{sd}|^2 l_{sd}+|{h}_{rd}|^2 l_{rd}+ 2\sqrt{l_{sd} l_{rd}} \Re(\rho {h}_{sd} {h}_{rd}^*)$, 
so:
\begin{align}
\prob\left(\adf^{c}\cap\bdf^{c}\right)\hspace{-1pt} &= \Ex_{\tilde{\Phi}}\left[  \prob\left\{|h_{sr}|^2 \geq \frac{TI_r}{\mu_3},Z\geq T I_d\biggl|{\tilde{\Phi}}\right\} \right],  \nonumber \\
&= \Ex_{\tilde{\Phi}} \left[ e^{-\frac{TI_r}{\mu_3}}\bar{F}_{V}(TI_d)\right], 
\end{align}
where $\bar{F}_{V}(\cdot)$ is the complementary cumulative distribution function of $V$ and $\mu_3$ is given by (\ref{eq:mu3}). The complementary cumulative distribution function  $Z$ is:
\begin{equation}
\bar{F}_{Z}(u) =\left\{ \begin{array}{cc}
\frac{\mu_2 e^{-u / \mu_2}-\mu_1 e^{- u / \mu_1}}{\mu_2-\mu_1} & \mu_1 \neq \mu_2 \\
(1+u/\mu_1) e^{-u/\mu_1} & \mu_1 = \mu_2
\end{array}\right.
 \label{eq:FV2}
\end{equation}
To see this, notice that $Z$ can be written as:
\begin{equation}
Z= q^H \left[\begin{array} {cc} l_{sd} & \sqrt{l_{sd} l_{rd}} \rho \\ \sqrt{l_{sd} l_{rd}} \rho^* & l_{rd} \end{array}\right] q := q^H Q q,
\end{equation}
where $q = [h_{sd},\ h_{rd}]^T$, is a zero-mean complex circularly symmetric Gaussian vector with identity covariance matrix and $Q$ is positive definite. Diagonalizing $Q$ we can write $Z = \mu_1 |w_1|^2 + \mu_2 |w_2|^2$, where $\mu_1$ and $\mu_2$  are the eigenvalues of $Q$, given  by (\ref{eq:mu1}) and (\ref{eq:mu2}), and $|w_1|^2$ and $|w_2|^2$ are unit mean exponential variables, the same as $|h_{sd}|^2$ and $|h_{rd}|^2$. Applying a straightforward change of variables from $\{|w_1|^2,|w_2|^2\}$ to $Z$ (\ref{eq:FV2}) is obtained. Notice that when $\rho=0$ and $||d-r||= D$ $\mu_1$ and $\mu_2$  coincide so $Z$ follows a Gamma distribution with $2$ degrees of freedom, which accounts for the case $\mu_1 = \mu_2$. 
Now, replacing that expression in (\ref{eq:PoutDF1}) and using the definition of the Laplace transform (\ref{eq:lapdef}) we obtain (\ref{eq:PoutDF2}). 
To obtain (\ref{eq:PoutDF3}) we replace (\ref{eq:FV2}) in (\ref{eq:PoutDF1}) and use the fact that:
\begin{equation}
\frac{d\Ex_{\tilde{\Phi}} \left[ e^{-(\omega_1 I_d + \omega_2 I_r)} \right]}{d\omega_1} = \Ex_{\tilde{\Phi}} \left[-I_d e^{-(\omega_1 I_d + \omega_2 I_r)} \right].
\end{equation}

\section{Proof of Theorem \ref{lem:OPbsingrf}} \label{proof:OPbsingrf}
Starting from (\ref{eq:UBfixedrelay}), we find the probability of the events $\adf$ and $\bdf$, by following the steps used to derive (\ref{eq:PoutDF1}):
\begin{align}
\hspace{-3mm}\prob \left\{\adf(R,\rho)\right\} &= 1-\mathcal{L}_{I_r} \left(T/(1-|\rho|^2)\right), \label{eq:PADF1}\\ 
\hspace{-3mm} \prob\left\{\bdf(R,\rho)\right\} &= 1-\frac{ \mu_2 \mathcal{L}_{I_d} \left(T/\mu_2\right) - \mu_1 \mathcal{L}_{I_d} \left(T/\mu_1\right)}{ \mu_2 - \mu_1}. \label{eq:PBDF1}
\end{align}
The Laplace transform can be found in Lemma \ref{lem:LapTransSimple} and setting $\omega_1$ or $\omega_2$ to zero as needed \cite{baccelli_aloha_2006}
$\mathcal{L}_{I_r}(\omega) =\mathcal{L}_{I_d}(\omega) =  \exp\{-\lambda C \omega^{2/\alpha}\},$
where $C$ is given by (\ref{eq:C}). Replacing (\ref{eq:PADF1}) and (\ref{eq:PBDF1}) in (\ref{eq:UBfixedrelay}) and using (\ref{eq:delta_def}) we finish the proof.
The proof of the second part follows directly from the fact that $e^{-u}=1-u+\Onot(u^2)$ and the definition of $\kappa(R)$ (\ref{eq:SNDR}).

\section{Proof of lemma \ref{lem:optrho}}
\label{Ap:rho_optimal}
The Laplace transform (\ref{eq:Lapgen1}) writes as:
\begin{equation}
\mathcal{L}_{I_d,I_r}(\omega_1,\omega_2) 
= 1-\lambda 
 \int_{\R^2} \hspace{-1mm} \left[ 1- \frac{1}{(1+\omega_1l(x,d))(1+\omega_2l(x,r))} \right] dx + O(\lambda^2) \hspace{-3mm}
\end{equation}
and hence, the OP (\ref{eq:first_out_fixed_realy}), using Theorem \ref{lem:pdffixed} can be written as:
\begin{equation*}
\hspace{-2mm}1 - \poutdf =O(\lambda^2)+ \lambda \int_{\R^2} \left\{\frac{\mu_3}{\mu_3+Tl(x,r)}
 \times \! \left[\frac{\mu_2^2  (\mu_2-\mu_1)^{-1}}{\mu_2+ Tl(x,d)}\!-\!\frac{\mu_1^2 (\mu_2-\mu_1)^{-1}}{\mu_1+ Tl(x,d)} \!-\! \frac{l_{sd}}{l_{sd}+Tl(x,d)}\right]\!\right\}dx. 
\end{equation*}
We now show that the integrand is decreasing in $|\rho|$. To do this, it is useful now to write $\mu_1 = a-b(\rho)$ and $\mu_2 = a+b(\rho)$, with $a=\frac{1}{2}(l_{sd}+l_{rd})$ and $b(\rho) = \frac{1}{2} \left[(l_{sd}-l_{rd})^2 - 4 l_{sd} l_{rd} |\rho|^2\right]^{\frac{1}{2}}.$
We can therefore write:
\begin{multline}
1 - \poutdf =O(\lambda^2)+ \lambda \int_{\R^2} \left\{\frac{\mu_3(|\rho|)}{\mu_3(|\rho|)+Tl(x,r)} \right.\\ \left.
 \times  \left[\frac{\varphi(b(|\rho|))+\varphi(-b(|\rho|))}{2} - \frac{l_{sd}}{l_{sd}+Tl(x,d)}\right]\right\}dx, \nonumber
\end{multline}
with:
$\varphi(u) =\frac{(a+u)^2}{u (a+u+Tl(x,d))}.$
By using that $\mu_3$ is decreasing in $|\rho|$ we see that:
\begin{equation}
\frac{\mu_3(|\rho|)}{\mu_3(|\rho|)+Tl(x,r)}
\end{equation}
is decreasing in $|\rho|$, and positive. To show that the rest is positive, we check that
\begin{equation*}
\frac{d (\varphi(b)\! + \! \varphi(-b))}{db} \! =  \!\frac{-4T^2l(x,d)^2 b}{(T^2 l(x,d)^2 +2 T l(x,d)a+ \!a^2 -b^2)^2} \! < \! 0,
\end{equation*}
and since $b$ is increasing in $|\rho|$, then $\varphi(b(|\rho|)) + \varphi(-b(|\rho|))$ is decreasing in $|\rho|$. We conclude taking $|\rho| = 1$ to check that:
\begin{equation*}
\frac{\varphi(b(1))+\varphi(-b(1))}{2} - \frac{l_{sd}}{l_{sd}+Tl(x,d)} >0.
\end{equation*}
\section{Proof of Theorem \ref{lem:psdffixed} } \label{proof:psdffixed}
In what follows we omit the dependence of the outage event on $(R,\epsilon)$. Using the concavity of the logarithm we can bound:
\begin{equation}
R_{SDF} \leq \tilde{R}_{SDF} = \mathcal{C} \left(\frac{|\hsd|^2l_{sd} + (1-\varepsilon) |\hrd|^2 l_{rd}}{I_d}\right)
\end{equation}
With this we have $\bsdf \supset \tilde{\mathcal{B}}_{SDF} = \{ \tilde{R}_{SDF} < R\}$ and:
\begin{align}
\mathcal{O}_{SDF} &\supset [\asdf^c \cap  \tilde{\mathcal{B}}_{SDF} ] \cup [\asdf \cap \adt],\\
&= [( \asdf^c \cap \tilde{\mathcal{B}}_{SDF}^c) \cup (\asdf \cap \adt^c) ]^c \label{eq:osdfbou}
\end{align}
In the second step we used that $\tilde{\mathcal{B}}_{SDF} \subset \adt$ and the union is disjoint. The rest of the proof follows along the lines of the proof in Appendix \ref{proof:pdffixed}:  the first event in (\ref{eq:osdfbou}) is the same as that of $\bdf$ by taking $\rho = 0$ and a path loss $(1-\varepsilon) l_{rd}$. The second term is the same as (\ref{eq:PoutDF1}) with $\rho = 0$ and $T= 2^{R/\epsilon}-1$.
\section{Proof of lemma \ref{lem:GapRates}} \label{proof:GapRates}
We use the CF rate given by (\ref{eq:rateCf2}), and write $R_1 \equiv R_1(0)$ and $R_2 \equiv R_2(0)$:
\begin{align*}
\!R_{CF} (0) \! - \!R_{CF}(\rho_N) &= \min\{\! R_1,R_2\} \! - \! \min\{\! R_1(\rho_N),R_2(\rho_N)\} \\
&\leq \max \{ R_1-R_1(\rho_N),R_2-R_2(\rho_N)\}.
\end{align*}
Now we use that:
\begin{equation}
R_1 - R_1(\rho_N) = \Cfunc\left(\frac{I_r}{n_c} (1-|\rho_N|^2) \right) -  \Cfunc\left(\frac{I_r}{n_c} \right) 
= \Cfunc \left(-\frac{|\rho_N|^2}{1+ \frac{n_c}{I_r} }\right) \leq 0. 
\end{equation}
On the other hand, defining: 
\begin{alignat}{2}
u = \frac{|\hsd|\sqrt{l_{sd}}}{\sqrt{I_d}} \hspace{5mm}&\hspace{5mm}
v = \frac{|\hsr|\sqrt{l_{sr}}}{\sqrt{I_r}}
\end{alignat}
we can rewrite and bound the rate $R_2(\rho_N)$ as:
\begin{align}
\hspace{-2mm}R_2(\rho_N ) &= \Cfunc\left(|u|^2 + \frac{|\rho_N|^2 |u|^2 + |v|^2 - 2 \Re\{\rho_N u v^*\}}{1 + \frac{n_c}{I_r} - |\rho_N|^2} \right), \nonumber\\
& \geq \Cfunc\left(|u|^2 + \frac{|\rho_N|^2 |u|^2 + |v|^2 - 2 |\rho_N| |u| |v|}{1 + \frac{n_c}{I_r} - |\rho_N|^2} \right). \label{eq:R2_r2}
\end{align}
Now we are interested in finding a lower bound of (\ref{eq:R2_r2}). Using standard analysis techniques it is straightforward to check that in the range $0 \leq \rho_N \leq 1$ this function has a single minimum. We have to consider three regimes:
\begin{itemize}
\item $|v| < |u|$: there is a minimum at $|\rho_N| = \frac{|v|}{|u|} < 1.$
\item  $|u| \leq |v| \leq (1+\frac{n_c}{I_r})|u|$: the function is decreasing $\rho_N$ so there $|\rho| = 1$ gives the smallest rate.
\item $(1+\frac{n_c}{I_r}) |u| < |v|$: in this case there is a minimum at:
\begin{equation}
|\rho_N| = \frac{(1+\frac{n_c}{I_r})|u|}{|v|} < 1.
\end{equation}
\end{itemize}
With this analysis, we can show that:
\begin{equation*}
R_2(\rho_N) \geq \begin{cases}
\Cfunc(|u|^2)  & \text{if $\frac{|v|}{|u|} < 1$,} \\
\Cfunc\left( |u|^2+ \frac{(|u|-|v|)^2I_r}{n_c}\right)& \text{if $1 \leq \frac{|v|}{|u|} \leq (1+\frac{n_c}{I_r})$,}\\
\Cfunc\left( \frac{|v|^2}{1+\frac{n_c}{I_r}} \right)&\text{if $(1+\frac{n_c}{I_r})< \frac{|v|}{|u|} $.}\\
\end{cases}
\end{equation*}
Using this fact we conclude by noting that:
\begin{equation*}
R_2(0) - R_2(\rho_N) \!\leq\! \begin{cases} 
\Cfunc\left(\frac{|v|^2}{|u|^2 \left(1+ \frac{n_c}{I_r}\right)} \right) \leq 1 & \text{if $\frac{|v|}{|u| (1+\frac{n_c}{I_r})} \leq 1$} \\ 
\Cfunc \left(\frac{|u|^2}{|v|^2 \left(1+\frac{n_c}{I_r}\right)}\right) \leq 1 & \text{if $\frac{|v|}{|u| (1+\frac{n_c}{I_r})} > 1$}, 
\end{cases}
\end{equation*}

\section{Proof of theorem \ref{lem:bouCFOP}} \label{proof:bouCFOP}
Since,
$R_{CF}(\rho_N,n_c) \geq R_{CF}(0,n_c) - 1$,
for an attempted rate $R$, we have:
\begin{align*}
\poutcf(R,n_c) &= \prob(R_{CF}(\rho_N,n_c) < R) \\
&\leq \prob(R_{CF}(0,n_c) - 1 < R).
\end{align*}
For the bound on the probability of $\mathcal{A}_{CF}$ we use that:
\begin{equation*}
\left\{\frac{|\hsr|^2 l_{sr}}{I_r+n_c} < T \right\}  =  \acf \cup \left( \acf^c \cap \left\{\frac{|\hsr|^2 l_{sr}}{I_r+n_c} < T \right\} \right). 
\end{equation*}
and that 
\begin{equation*}
 \acf^c \cap \left\{\frac{|\hsr|^2 l_{sr}}{I_r+n_c} < T \right\} \subseteq 
 \bigcup_{i = 0}^{N-1} \hspace{-.7mm} \left\{ \hspace{-.7mm} \frac{n}{N}T \hspace{-.1mm} \leq  \hspace{-.1mm} \frac{|\hsr|^2 l_{sr}}{I_r+n_c} \hspace{-.1mm} \!<\! \hspace{-.1mm} \frac{n+1}{N} T \right. \hspace{-0.6mm}, 
 \left. \frac{|\hsd|^2 l_{sd}}{I_d} \hspace{-.1mm} \geq \hspace{-.1mm}\frac{N-n}{N} T \right\} \hspace{-0.9mm},
\end{equation*}
with $N$ a natural number. The union in the previous equation is a disjoint coverage of the event on the left side, so we have:
\begin{equation*}
\prob(\mathcal{A}_{CF}) \leq
\prob\left(\frac{|\hsr|^2 l_{sr}}{I_r+n_c} < T \right) - 
 \sum_{n=0}^{N-1} \hspace{-.7mm} \prob \hspace{-.7mm} \left(\frac{n}{N}T \hspace{-.1mm}\leq  \hspace{-.1mm} \frac{|\hsr|^2 l_{sr}}{I_r+n_c} \hspace{-.1mm}  <  \hspace{-.1mm} \frac{n+1}{N} T \right. \hspace{-0.6mm}\!,\! 
\left. \frac{|\hsd|^2 l_{sd}}{I_d} \hspace{-.1mm} \geq \hspace{-.1mm}\frac{N-n}{N} T\right) \hspace{-0.9mm}.
\end{equation*}
Now we can condition on the point process and using that the fading coefficients are independent
we can write the probabilities in terms of the Laplace transform of the interferences.
For the other event, since $n_c > 0 $ we have:
\begin{align}
\hspace{-2mm}\!\bar{\mathcal{A}}_{CF}(R+1,n_c,0) &\subseteq \bar{\mathcal{A}}_{CF}(R+1,0,0) \nonumber \\
&= \left\{\!\frac{1}{T}\! \left(\!\frac{|\hsr|^2l_{sr}}{I_r} + \frac{|\hsd|^2 l_{sd}}{I_d}\!\right) \geq 1 \right\}\!.\! \label{eq:AcfBou1}
\end{align}
Noticing that:
\begin{equation}
\bcf(n_c,0)= \left\{ n_c < \frac{I_r I_d}{|\hrd|^2}  \left( \hspace{-1mm}\frac{|\hsd|^2 }{I_d}+ \frac{|\hsr|^2}{I_r}+1\right)\hspace{-1mm} \right\},\hspace{-1mm}\label{eq:BcfNc0-1}
\end{equation}
we see that we can use (\ref{eq:BcfNc0-1}) with (\ref{eq:AcfBou1})  to bound:
\begin{align}
\bar{\mathcal{A}}_{CF}(R+1,n_c,0) \cap \bcf(n_c,0) 
&\subseteq \bar{\mathcal{A}}_{CF}(R+1,0,0) \cap \bcf(n_c,0) \\
&\subseteq \left\{\frac{ |\hsr|^2 l_{sr}  I_d + |\hsd|^2 l_{sd} I_r}{n_c l_{rd} |\hat{h}_{rd}|^2} > \frac{T}{1+T}\right\}.
\end{align}
We therefore have:
\begin{align*}
\prob\left(\bar{\mathcal{A}}_{CF}(R+1,n_c,0) \cap \bcf(n_c,0) \right)  
&\leq \prob \left( \frac{ |\hsr|^2 l_{sr}  I_d + |\hsd|^2 l_{sd} I_r}{n_c l_{rd} |\hrd|^2} > \frac{T}{1+T} \right) \\
&=   1 \hspace{-0.3mm} - \hspace{-0.3mm} \Ex \hspace{-0.4mm}\left[ \mathcal{L}_{I_d,I_r}\hspace{-0.9mm} \left( \hspace{-0.4mm}\frac{(1+T)l_{sr} |\hsr|^2}{T n_c l_{rd}},\frac{(1+T)l_{sd} |\hsd|^2}{T n_c l_{rd}} \right) \hspace{-0.5mm} \right], \hspace{-3mm}
\end{align*}
where the expectation is over $|\hsr|^2$ and $|\hsd|^2$. To obtain this last expression we first condition on $\hsr$, $\hsd$ and the point process, and evaluate the probability with respect to $\hrd$. We then take the expectation with respect to the point process to obtain the joint Laplace transform and finally the expectation with respect to the fading coefficients. All these random elements are independent. By using this bound we avoid working with the
product of the interference at the relay and the destination, which complicates the evaluation of $\prob(\mathcal{B}_{CF})$
significantly. In addition, when all the distances remain fixed and for sufficiently small $\lambda$, the product term will be small and its contribution will not be significant in comparison with the other terms. 
Now observing the expression of the Laplace transform given in lemma \ref{lem:LapTransSimple} we see that we can lower bound the joint Laplace Transform by removing the function (\ref{eq:integral_hard}), which is equivalent to assuming that the interferences are independent and splits the joint Laplace transform into the product of the transforms of the separate interferences.
\bibliographystyle{IEEEtran}
\bibliography{IEEEabrv,relaygeo}

\end{document}